\documentclass[12pt,superscriptaddress]{iopart}

\usepackage{epsfig}
\usepackage{graphicx}
\usepackage{dcolumn}
\usepackage{bm} 
\usepackage{epstopdf}

\newcommand{\beq}{\begin{equation}}
\newcommand{\eeq}{\end{equation}}
\newcommand{\beqa}{\begin{eqnarray}}
\newcommand{\eeqa}{\end{eqnarray}}

\def\jpb#1{{ J.\ Phys.\ B} {\bf#1}}

\def\jpg#1{{ J.\ Phys.\ G} {\bf#1}}

\def\natphot#1{{ Nat.\ Photonics} {\bf#1}}

\def\pra#1{{ Phys.\ Rev. A\/} {\bf#1}}

\def\prl#1{{ Phys.\ Rev.\ Lett.} {\bf#1}}

\def\rmp#1{{ Rev.\ Mod.\ Phys.} {\bf#1}}
\def\sci#1{{ Science} {\bf#1}}
\def\nat#1{{ Nature} {\bf#1}}

\begin{document}

\title[Alpha decay in intense laser fields: Calculations using realistic nuclear potentials]{Alpha decay in intense laser fields: Calculations using realistic nuclear potentials}

\author{Jintao Qi$^1$, Tao Li$^2$, Ruihua Xu$^1$, Libin Fu$^1$, and Xu Wang$^{1,*}$}

\address{$^1$ Graduate School, China Academy of Engineering Physics, Beijing 100193, China}
\address{$^2$ Beijing Computational Science Research Center, Beijing 100193, China}
\ead{$^*$xwang@gscaep.ac.cn}

\begin{abstract}
We calculate the effect of intense laser fields on nuclear alpha decay processes, using realistic and quantitative nuclear potentials. We show that alpha decay rates can indeed be modified by strong laser fields to some finite extent. We also predict that alpha decays with lower decay energies are relatively easier to be modified than those with higher decay energies, due to longer tunneling paths for the laser field to act on. Furthermore, we predict that modifications to angle-resolved penetrability are easier to achieve than modifications to angle-integrated penetrability.

\end{abstract}

\maketitle

\section{Introduction}

The past few decades witness rapid advancements in intense laser technologies. The chirped pulse amplification technique \cite{CPA} enables table-top Ti:sapphire lasers to have intensities exceeding one atomic unit ($3.5\times10^{16}$ W/cm$^2$), opening the door to the rich area of strong-field atomic, molecular, and optical physics with novel nonperturbative phenomena such as multiphoton and above-threshold ionization\cite{Agostini-79,Paulus-94}, high harmonic generation \cite{McPherson-87,Ferray-88}, nonsequential double and multiple ionization \cite{Walker-94,Palaniyappan-05}, attosecond physics \cite{Krausz-RMP-09,Zhao2012,Li2017,Gaumnitz2017}, etc. 

Even higher intensities can be achieved by larger laser systems of different kinds, for example, X-ray free electron lasers (XFELs) and the under-constructing extreme light infrastructure (ELI) of Europe. XFELs can be focused to reach peak intensities on the order of $10^{20}$ W/cm$^2$ \cite{Bucksbaum-2015}, and ELI is designed to reach peak intensities above $10^{23}$ W/cm$^2$ \cite{Ursescu-2013,Ur-2015}. The laser electric field corresponding to such an intensity is comparable to the Coulomb field from the bare nucleus at a distance of about 10 fm. Direct influence on the nucleus may be possible from such an intense laser field. In fact, one of the major scientific goals of the ELI facility is to study laser-driven nuclear physics. 

Direct light-nucleus interaction with much weaker light intensities has been realized using synchrotron radiations on the M\"ossbauer $^{57}$Fe system. Using a grazing-incidence X-ray diffraction technique and a planar $^{57}$Fe cavity, collective quantum optical effects have been demonstrated with photon energy 14.4 keV, such as single-photon superradiance \cite{Rohlsberger-2010}, electromagenetically induced transparency \cite{Rohlsberger-2012}, spontaneously generated coherence \cite{Heeg-2013}, Rabi oscillation \cite{Haber-2017}, etc. On the other hand, the nuclei, as the media of X-ray pulse propagation, can be used to modify the properties of the X-ray pulse, such as the pulse shape \cite{Vagizov-2014} and the group velocity \cite{Heeg-2015}. Theoretical proposals have also been made on single-photon entanglement \cite{Palffy-2009}, single-photon storage and phase modulation \cite{Liao-2012}, nuclear battery using isometric transition \cite{Liao-2012-2}, etc. 

Non-resonant effects of intense laser fields on nuclear systems have also been reported in the literature. Among them possible influence of intense laser fields on nuclear alpha decay has received attention \cite{Misicu-2013,Misicu-2016,Delion-2017}. Widely accepted as a quantum tunneling process \cite{Gamow-1928}, alpha decay is expected to be modified in the presence of a strong laser field through modifying the potential barrier, on which quantum tunneling depends very sensitively. Indeed, existing works all predict such modifications. 

To what degree an intense laser field, currently available or to be available in the forthcoming years, can influence alpha decay? This quantitative question, however, remains unanswered. Mi\c sicu and Rizea numerically solve a time-dependent Schr\"odinger equation using a one-dimensional (1D) model nuclear potential \cite{Misicu-2013,Misicu-2016}. They focus on obtaining qualitative understandings instead of quantitative evaluations. Delion and Ghinescu \cite{Delion-2017} adopt a Kramers-Henneberger (KH) approximation \cite{Kramers-1956,Henneberger-1968} to describe the laser-nucleus interaction. However, as will be explained in detail in the following section, the KH approximation is not valid to describe the laser-nucleus interaction. This explains why unreasonable predictions are made in \cite{Delion-2017} that the laser field greatly suppresses (by orders of magnitude) alpha decay along the polarization direction, where the laser electric field is the strongest, and greatly enhances (by orders of magnitude) alpha decay along the perpendicular direction, where no laser electric field is present. 

The goal of the current article is to study the effect of intense laser fields on nuclear alpha decay quantitatively. To achieve this goal we need to start with a realistic and quantitative alpha-nucleus potential. In this work we use the potentials proposed by Igo \cite{Igo-1958}, which has a simple analytical form and can be applied to a variety of nuclei. These potentials were obtained by fitting to alpha-nucleus scattering data. Our numerical results show that alpha decay can indeed be modified by strong external laser fields, to some small but finite extent. For example, with a laser intensity of $10^{24}$ W/cm$^2$, which is expected to be achievable in the forthcoming years with ELI, a modification of 0.1\% to the alpha particle penetrability or the nuclear half life is predicted. Besides, the alpha decay is modified along the laser polarization direction, as would be expected reasonably. A somewhat surprising result is that alpha decays with lower decay energies are relatively easier to be modified by external laser fields. This is due to longer tunneling paths under the potential barrier for the laser field to act on. We also explain that modifications to angle-resolved penetrability are easier to achieve than modifications to angle-integrated penetrability. The former modifications depend linearly on the laser electric field strength while the latter modifications depend quadratically on the laser electric field strength.

This article is organized as follows. In Section 2 we will introduce the methods that we use in our calculations. That includes the detailed form of the alpha-nucleus potentials, the form of the laser-nucleus interaction, and the method to calculate the alpha particle penetrability. Numerical results, analyses, and discussions are given in Section 3. A conclusion is given in Section 4.

\section{Method}

\subsection{The alpha-nucleus potential}

The potential energy felt by the alpha particle from the residue (daughter) nucleus can be written as
\beq \label{e.Vr}
V(r) = V_N(r) + V_C(r),
\eeq
where $r$ is the distance between the alpha particle and the daughter nucleus, $V_N(r)$ is a short-range nuclear potential and $V_C(r) = 2Z/r$ is the Coulomb repulsive potential. $Z$ is the charge of the daughter nucleus. 

Igo proposed a quantitative yet simple alpha-nucleus potential \cite{Igo-1958} by fitting to alpha-nucleus scattering data 
\beq
V_N(r) = -1100 \exp \left\{ -\left[ \frac{r-1.17A^{1/3}}{0.574} \right] \right\} \ \textrm{MeV},
\eeq
where $r$ is in units of fm (1 fm = $10^{-15}$ m) and $A$ is the mass number of the daughter nucleus. The potential is given in MeV. 

\begin{figure} 
\includegraphics[width=5cm]{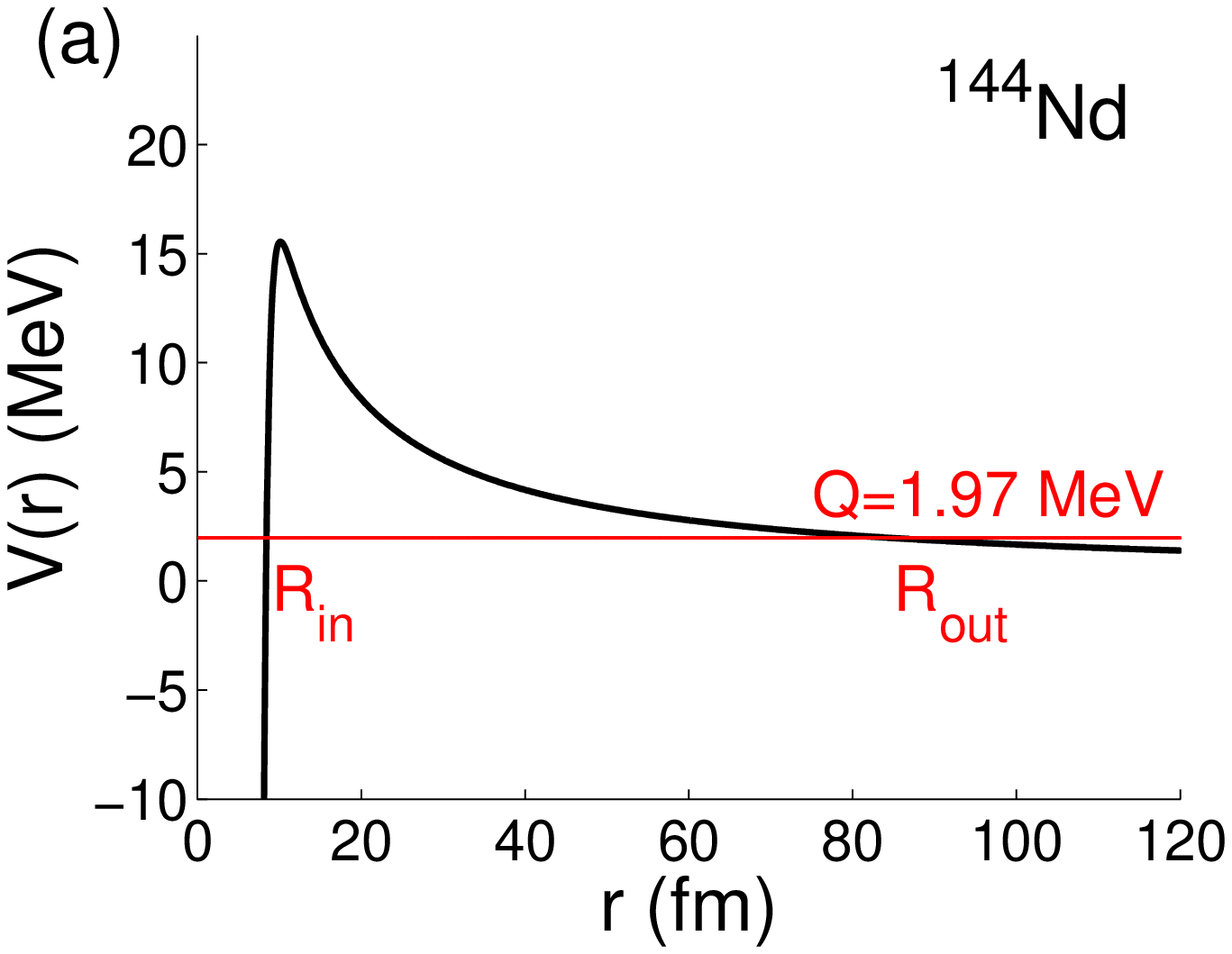}
\includegraphics[width=5cm]{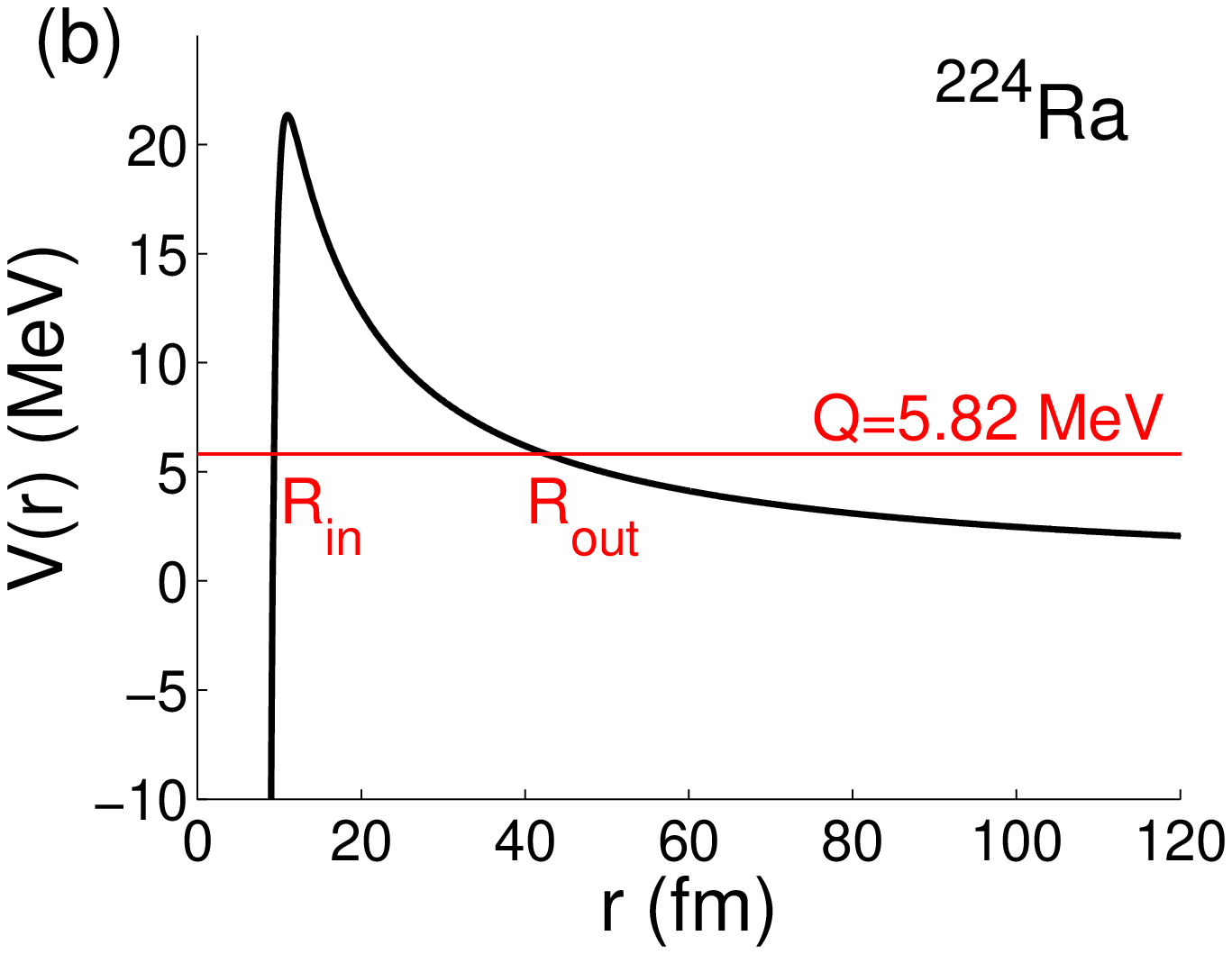}
\includegraphics[width=5cm]{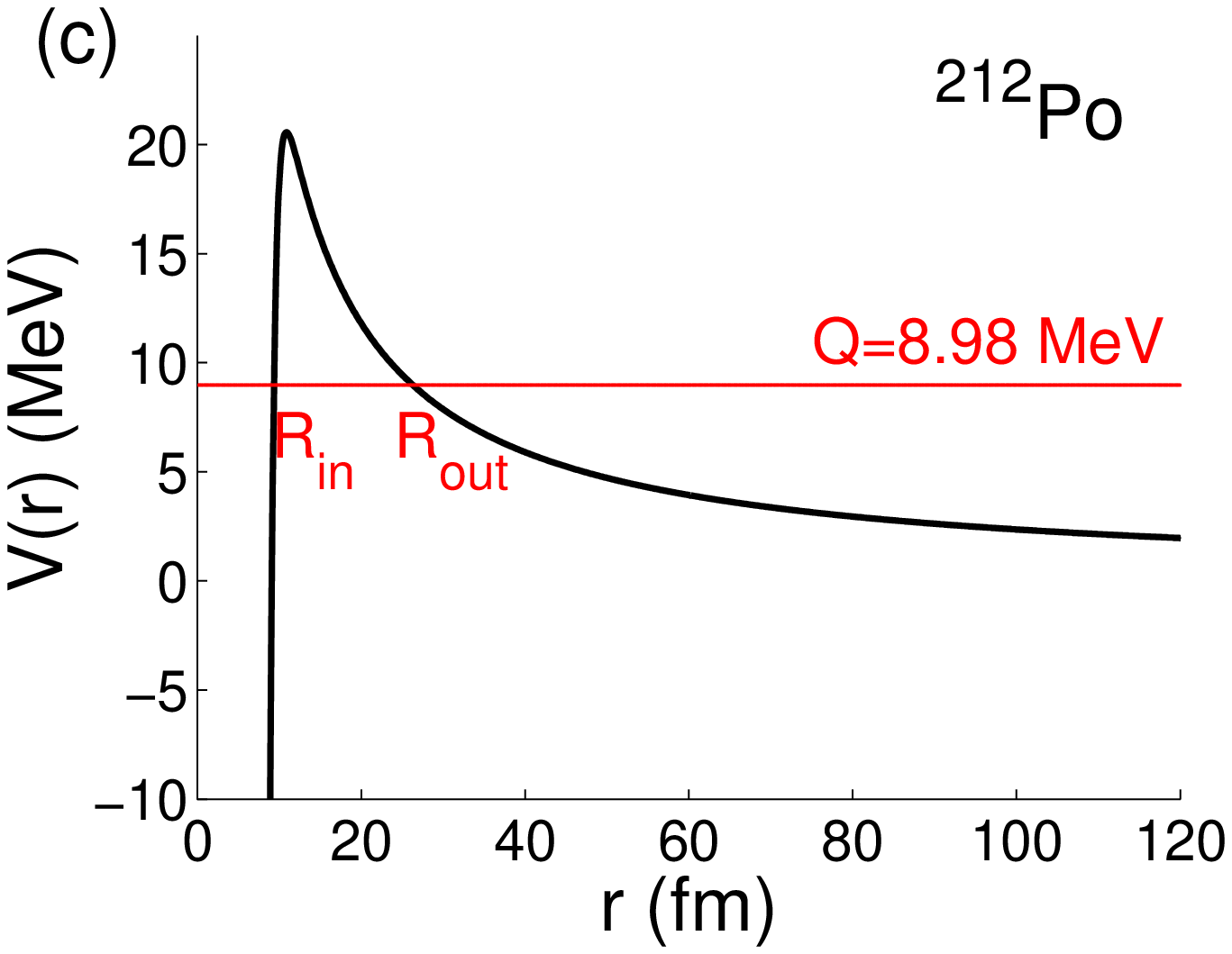}
\caption{Potentials between the alpha particle and the daughter nucleus, including the Igo alpha-nucleus potential the the Coulomb potential, for three nuclear elements (a) $^{144}$Nd (b) $^{224}$Ra and (c) $^{212}$Po. The red horizontal line in each panel is the corresponding alpha decay energy $Q$, the intersections of which with the potential energy curve give the tunneling entrance point $R_{in}$ and the tunneling exit point $R_{out}$.}  \label{f.potential}
\end{figure}

Figure \ref{f.potential} shows the potential $V(r)$ for three representative alpha-decay elements, namely, $^{144}$Nd, $^{224}$Ra, and $^{212}$Po. The decay energy $Q$ for the three elements are 1.97, 5.82, and 8.98 MeV, respectively. Typical alpha-decay energies range within 1 MeV to 10 MeV, so the three elements chosen represent low-, medium-, and high-energy decays.

The decay energy $Q$ contains three parts: the kinetic energy of the alpha particle, the recoil energy of the daughter nucleus, and the electron-screening correction due to energy loss of the alpha particle flying through the electron cloud of the atom. The electron-screening correction is much smaller than the other two parts. $Q$ can be written as
\beq \label{e.Q}
Q = \frac{A+4}{A} E_{\alpha} + \left[ 65.3(Z+2)^{7/5} - 80.0 (Z+2)^{2/5} \right]\times10^{-6} \ \textrm{MeV},
\eeq
where $E_{\alpha}$ is the kinetic energy of the alpha particle, and $Z$($A$) is the charge (mass number) of the daughter nucleus. The first term on the right hand side includes the energy of the alpha particle and the recoil energy of the daughter nucleus, and the second term is the electron-screening energy suggested by Perlman and Rasmussen \cite{Perlman-1957}.

\subsection{The laser-nucleus interaction}

The interaction between the laser electric field and the nucleus is given in the length gauge as \cite{Misicu-2013}
\beq \label{e.VI}
V_I (\vec{r},t) = -Z_{eff} \vec{r} \cdot \vec{\varepsilon} (t) = -Z_{eff} r \varepsilon(t) \cos\theta
\eeq
where $\theta$ is the angle between $\vec{r}$ and $\vec{\varepsilon} (t)$, and $Z_{eff} = (2A-4Z)/(A+4)$ is an effective charge. This effective charge indicates the tendency of the laser electric field separating the alpha particle and the daughter nucleus. One sees that if $Z/A=1/2$, then $Z_{eff} = 0$. That is, if the daughter nucleus has the same charge-to-mass ratio as the alpha particle, then the daughter nucleus and the alpha particle will move in concert in the laser field and the laser electric field does not have an effect of separating the two. For the three nuclear elements shown in Fig. \ref{f.potential}, $Z_{eff} = 0.33$ for $^{144}$Nd, 0.43 for $^{224}$Ra, and 0.42 for $^{212}$Po. 

The neglect of the magnetic part of the laser field is justified by the fact that the alpha particle moves much slower than the speed of light. Assuming a kinetic energy of 10 MeV, one gets an alpha particle speed of $2.2\times10^7$ m/s, about 7\% the speed of light. Therefore the effect of the magnetic field on the alpha particle is expected to be much smaller than that of the electric field.

\subsection{The quasistatic approximation}

The size of a typical nucleus is on the order of 1 fm. From a classical picture, the alpha particle oscillates back and forth within the nucleus. The frequency of this oscillation can be estimated to be $\sim 2\times10^7 \textrm{m}\cdot\textrm{s}^{-1} / 2 \textrm{ fm} = 10^{22}$ Hz. Each time the alpha particle hits the potential wall, it has a small chance (which is called the penetrability) of tunneling out. If it does, we may estimate how much time the alpha particle needs to tunnel through the potential barrier. Referring Fig. \ref{f.potential}, the length of the potential barrier for the alpha particle to tunnel through is on the order of 10 fm. So the alpha particle needs about $10^{-21}$ s to travel through the potential barrier. This time is much smaller than an optical cycle of strong lasers. For the 800 nm near-infrared laser of ELI, one optical cycle is $2.6\times10^{-15}$ s. For 10 keV X-ray lasers, one optical cycle is $4\times10^{-19}$ s. Therefore during the time that the alpha particle penetrates through the potential barrier, the change of the laser field is negligible and the laser field can be viewed as static. This is the quasistatic approximation. In strong-field atomic physics, such approximation is routinely used describing tunneling ionization of atoms \cite{ADK,Brabec-1996,Chen-2000,Yudin-2001}.

It is obvious that the Kramers-Henneberger approximation \cite{Kramers-1956,Henneberger-1968} is not valid. The KH approximation says that when the laser frequency is much higher than the particle oscillating frequency, the particle responses dominantly to the cycle-averaged laser field value (like our eyes' response to light). This high-frequency condition of validity for the KH approximation is well known in the literature \cite{Gavrila-1984,Pont-1990}. Applying the KH approximation to the laser-assisted alpha decay process has led to unreasonable predictions by Delion and Ghinescu \cite{Delion-2017}, as mentioned previously in the Introduction.

\subsection{The penetrability of the alpha particle}

Based on the quasistatic approximation, the penetrability of the alpha particle through the potential barrier can be calculated using the Wentzel-Kramers-Brillouin (WKB) method as
\beq \label{e.penetrability}
P(\theta,t) = \exp\left( -\frac{2}{\hbar}\int_{R_{in}}^{R_{out}} \sqrt{2\mu [V(r) - Q + V_I(r,\theta,t)]} dr \right),
\eeq 
where $V(r)$ and $V_I(r,\theta,t)$ are given in Eqs. (\ref{e.Vr}) and (\ref{e.VI}), respectively. The laser polarization is assumed to be along the z axis and $\theta$ denotes the direction of alpha emission, with respect to the +z axis. Understanding from the classical picture, the alpha particle oscillates back and forth inside the nucleus, and every time it hits the potential wall, it has a probability of $P(\theta,t)$ to tunnel out. 

In this article we mainly look into the relative change of the penetrability induced by the laser field. The relative change of the penetrability is defined as
\beq
\Delta = \frac{P(\varepsilon)-P(\varepsilon=0)}{P(\varepsilon=0)},
\eeq
where $\varepsilon$ is the laser field strength. $\Delta$ is also understood as a function of the emission angle $\theta$ and time $t$.

\section{Results and discussions}

\subsection{Laser-induced modifications to the penetrability}

\begin{figure} 
\includegraphics[width=5cm]{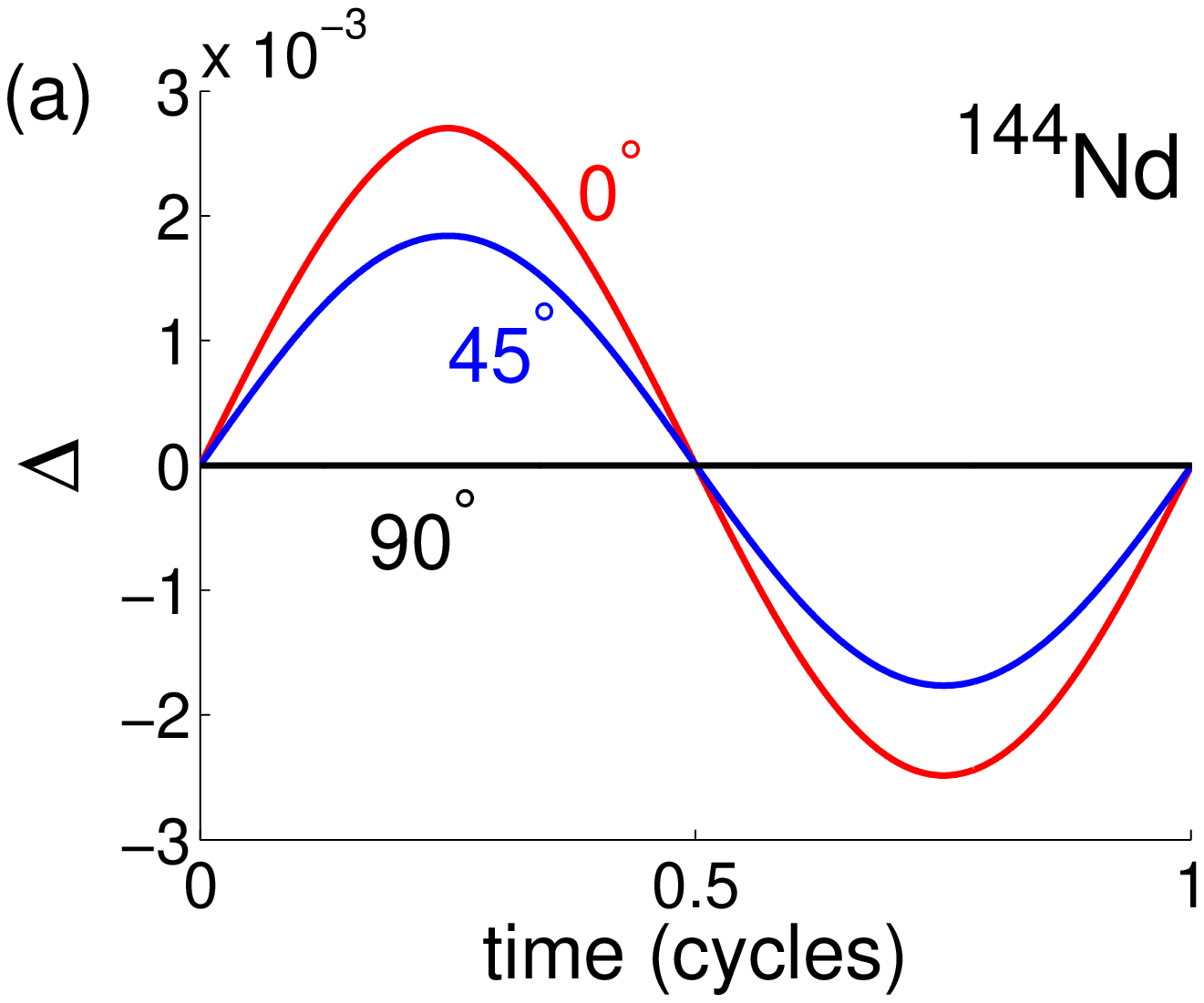}
\includegraphics[width=5cm]{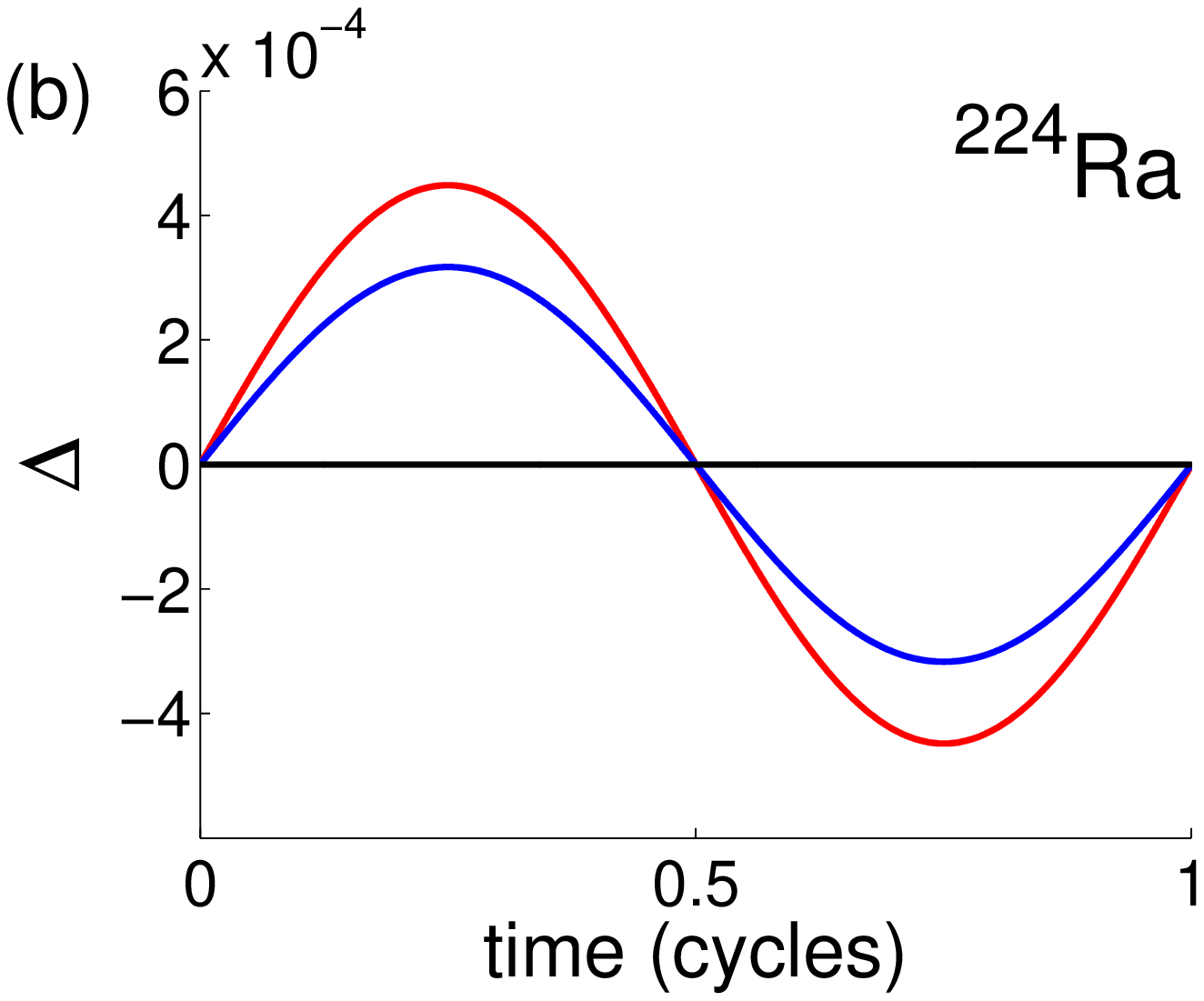}
\includegraphics[width=5cm]{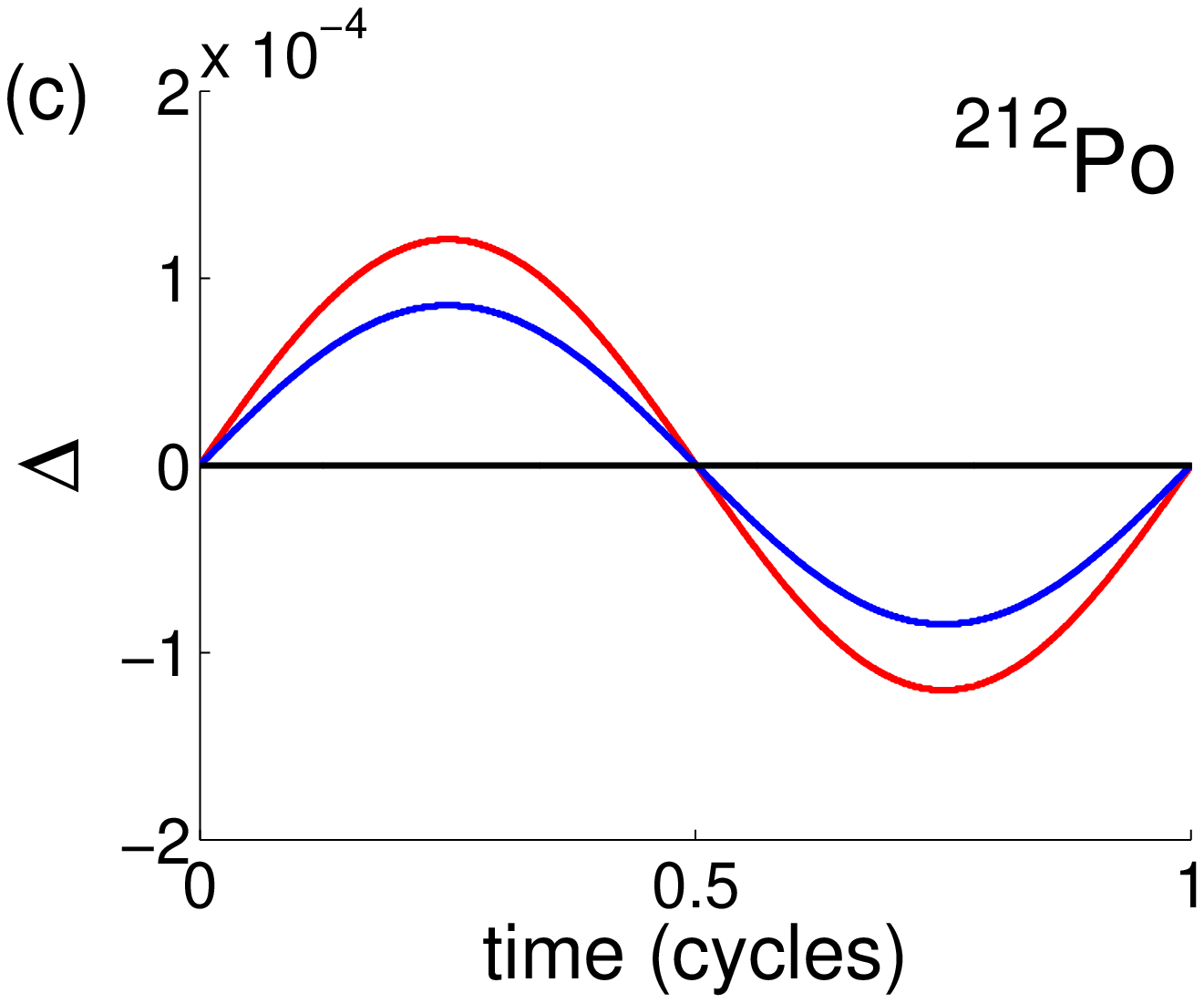} 
\caption{Time-dependent modifications to the alpha particle penetrability, seen from three spatial angles, namely, $\theta = 0^\circ$ (red), $45^\circ$ (blue), and $90^\circ$ (black). The same three nuclear elements are used as in Fig. \ref{f.potential}. The peak intensity used is $10^{24}$ W/cm$^2$ for all the three elements.}  \label{f.delta_time}
\end{figure}

First we show that the penetrability of the alpha particle can indeed be modified by strong external laser fields. Figure \ref{f.delta_time} shows time-dependent modifications to the penetrability seen from three spatial angles with respect to the +z direction, namely, $\theta = 0^\circ$ (red curves), $\theta = 45^\circ$ (blue curves), and $\theta = 90^\circ$ (black curves). The same three nuclear elements are used as in Fig. \ref{f.potential}. The modifications are strongest along $\theta = 0^\circ$, i.e., when the alpha emission direction is parallel to the laser polarization direction. No modifications are seen along $\theta = 90^\circ$, when the emission direction is perpendicular to the laser polarization direction. 

A peak intensity of $10^{24}$ W/cm$^2$ is used for all the three nuclear elements. This intensity is expected to be achieved by ELI in the forthcoming years. One sees that modifications to the alpha penetrability are on the order of 0.1\% for $^{144}$Nd and of 0.01\% for $^{224}$Ra or $^{212}$Po. The same amount of modifications are made to the nuclear half lives.

It may seem unexpected at first that $^{144}$Nd, with a lower decay energy than $^{224}$Ra and $^{212}$Po, is relatively easier to be modified by external laser fields. This is a consequence of the tunneling mechanism. $^{144}$Nd has a longer tunneling path for the laser field to act on, as shown in Fig. \ref{f.potential}, and the potential from the laser electric field is proportional to this path length.

\subsection{Laser potential as a perturbation to the alpha-nucleus potential}

Compared to the potential energy between the alpha particle and the daughter nucleus, the laser potential has much smaller magnitudes, even with an intensity of $10^{24}$ W/cm$^2$. We can gain insights into the laser-modification process by treating the laser potential as a perturbation to the alpha-nucleus potential.

Let us start from the penetrability exponential given in Eq. (\ref{e.penetrability}) and write it in the following form
\beqa
P(\theta,t) &=& \exp\left\{ -\frac{2\sqrt{2\mu}}{\hbar}\int_{R_{in}}^{R_{out}} dr \sqrt{V_0 + V_I} \right\} \\
  &=& \exp\left\{ -\frac{2\sqrt{2\mu}}{\hbar}\int_{R_{in}}^{R_{out}} dr \sqrt{V_0} \sqrt{1 + \frac{V_I}{V_0} } \right\}
\eeqa
where for convenience $V_0(r) \equiv V(r)-Q$. By assuming $|V_I|\ll |V_0|$, we have the following Taylor expansion
\beqa
P(\theta,t) &=& \exp\left\{ -\frac{2\sqrt{2\mu}}{\hbar}\int_{R_{in}}^{R_{out}} dr \sqrt{V_0} \left[ 1 +                      \frac{V_I}{2V_0} - \frac{V_I^2}{8V_0^2} + ...\right] \right\} \\
&\approx& \exp\left( \gamma^{(0)} + \gamma^{(1)} + \gamma^{(2)} \right) \\
&=& \exp\left( \gamma^{(0)} \right) \exp\left( \gamma^{(1)} + \gamma^{(2)} \right) \\
&\approx& P(\varepsilon=0, \theta, t) \left( 1 + \gamma^{(1)} + \gamma^{(2)} \right) 
\eeqa
where $\gamma^{(0)}$,$\gamma^{(1)}$,and $\gamma^{(2)}$ are defined as
\beqa
\gamma^{(0)} &=& -\frac{2\sqrt{2\mu}}{\hbar}\int_{R_{in}}^{R_{out}} dr \sqrt{V_0(r)}  \\
\gamma^{(1)} &=& \varepsilon(t) \frac{\sqrt{2\mu} Z_{eff} \cos\theta}{\hbar} \int_{R_{in}}^{R_{out}}  \frac{rdr}{\sqrt{V_0(r)}}  \\
\gamma^{(2)} &=&  \varepsilon^2(t) \frac{\sqrt{2\mu} Z_{eff}^2 \cos^2\theta}{4\hbar}\int_{R_{in}}^{R_{out}} \frac{r^2 dr}{V_0^{3/2}(r)}  \label{e.gamma2}
\eeqa
Note that $\gamma^{(0)}$ is independent of the laser electric field, $\gamma^{(1)}$ is proportional to $\varepsilon(t)$, and $\gamma^{(2)}$ is proportional to $\varepsilon^2(t)$.

\begin{figure} 
\includegraphics[width=5cm]{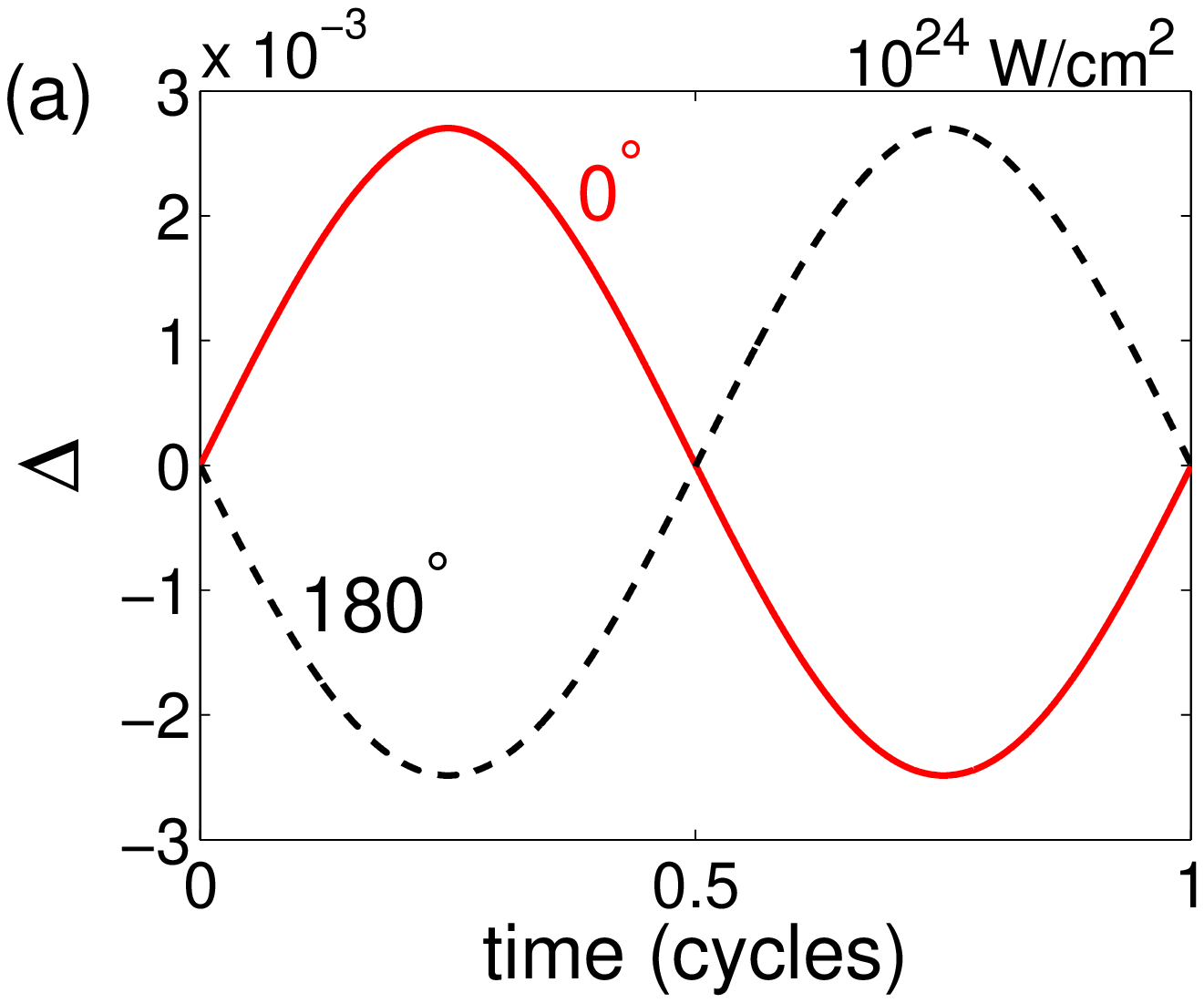}
\includegraphics[width=5cm]{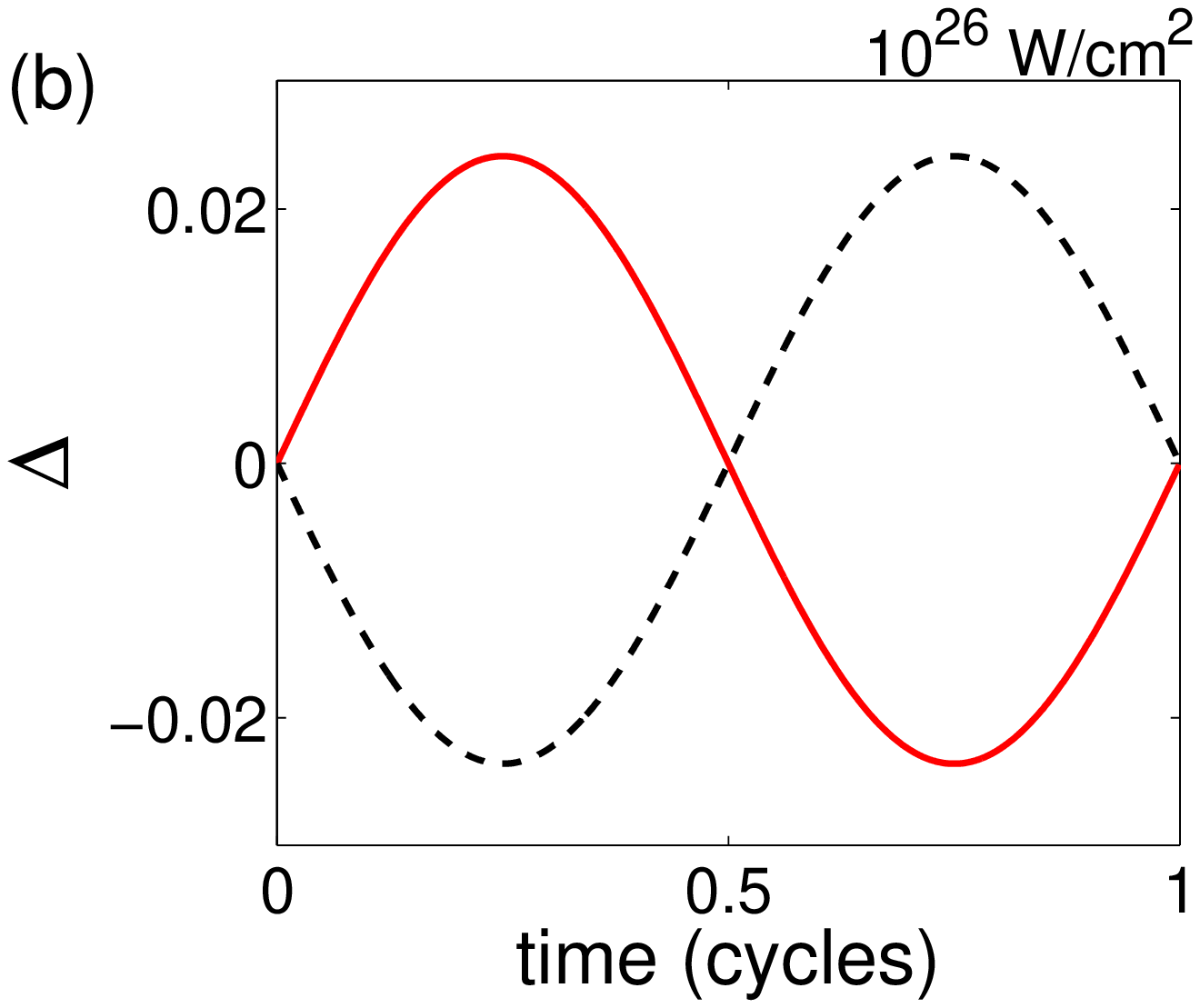}
\includegraphics[width=5cm]{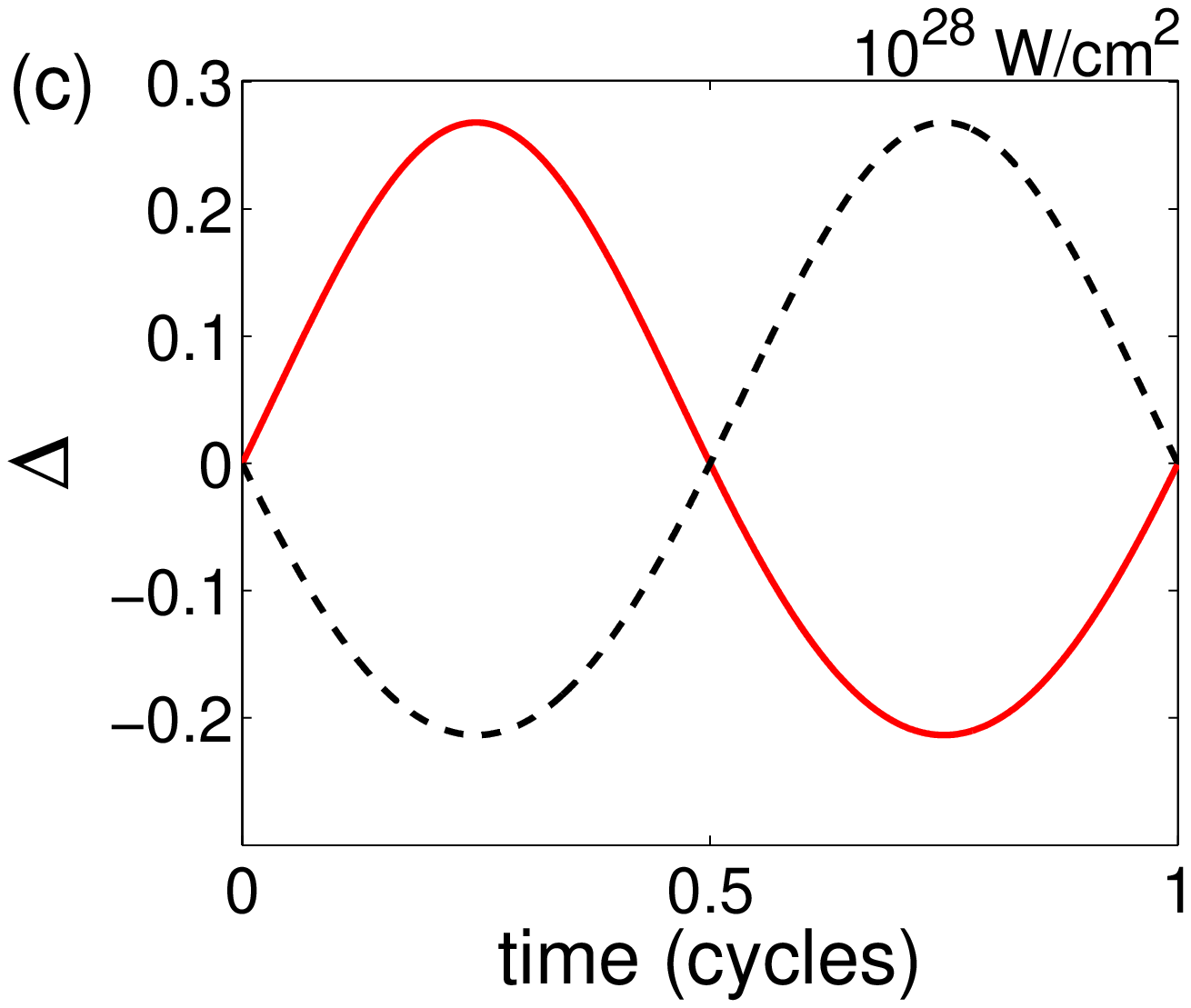} 
\caption{Relative changes of the penetrability $P$ seen from $\theta=0^\circ$ and from $\theta=180^\circ$, for three different laser intensities, namely, $10^{24}$ W/cm$^2$ (a), $10^{26}$ W/cm$^2$ (b), and $10^{28}$ W/cm$^2$ (c). The nuclear element used here is $^{144}$Nd. Clear positive-negative asymmetry between $0^\circ$ and $180^\circ$ can be seen in (c).}   \label{f.0_180}
\end{figure}

\subsection{$0^\circ$ versus $180^\circ$}

When there is no laser field, alpha emission to the direction $\theta=0^\circ$ is equivalent to $\theta=180^\circ$. When the laser electric field is on and pointing to $0^\circ$, then the penetrability to the same direction increases, and at the same time the penetrability to the opposite direction ($180^\circ$) decreases.

For intensities with which $\gamma^{(2)}$ is negligible, the response of the penetrability to the laser electric field is linear. This means that the amount that $P$ increases along $0^\circ$ is equal to the amount that $P$ decreases along $180^\circ$. The same argument can be made to other emission directions as well. Then there will be no net gain in the total alpha decay rate integrating over all emission directions. Only for higher intensities with which $\gamma^{(2)}$ is not negligible, does the total alpha decay rate increase. This can be seen from Eq. (\ref{e.gamma2}) that $\gamma^{(2)}$ is always positive so both $0^\circ$ and $180^\circ$ contribute positively to the decay rate. Therefore modifying the angle-integrated total decay rate requires much higher laser intensities than modifying the angle-resolved decay rates. The former is a second-order process in laser field strength, whereas the latter is a first-order process in laser field strength.

Figure \ref{f.0_180} shows the comparison between the modification to the alpha decay rate seen from $0^\circ$ and from $180^\circ$, for three different intensities, namely, $10^{24}$ W/cm$^2$, $10^{26}$ W/cm$^2$, and $10^{28}$ W/cm$^2$. The nuclear element used is $^{144}$Nd. One can see that for the lower two intensities, $0^\circ$ and $180^\circ$ look quite symmetric, at least from naked eyes, indicating small $\gamma^{(2)}$ values. For the relatively high intensity shown in panel (c), obvious positive-negative asymmetry can be seen, due to appreciable $\gamma^{(2)}$ values with this intensity.

\subsection{Angle-resolved versus angle-integrated modifications}

\begin{figure} 
\includegraphics[width=5cm]{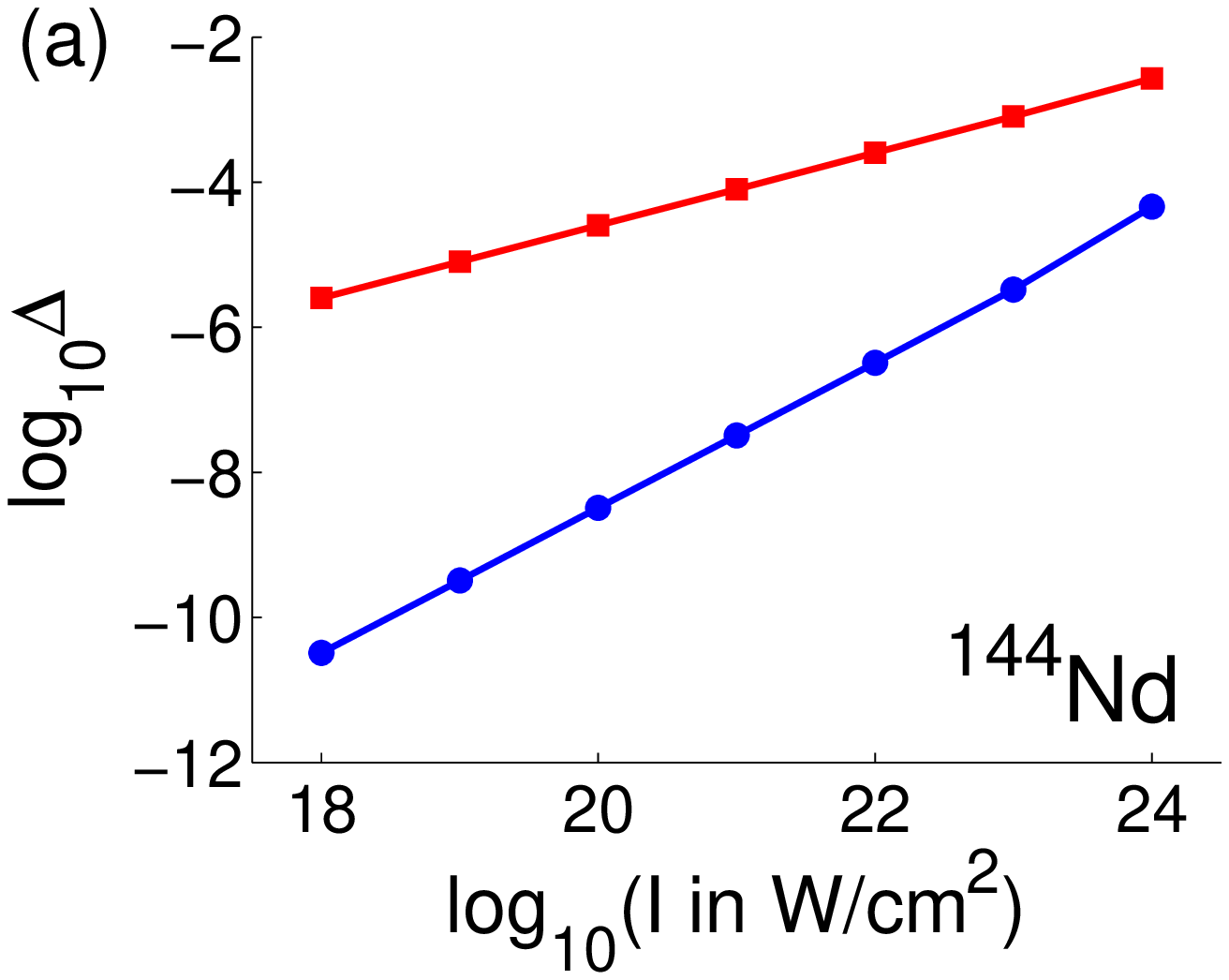}
\includegraphics[width=5cm]{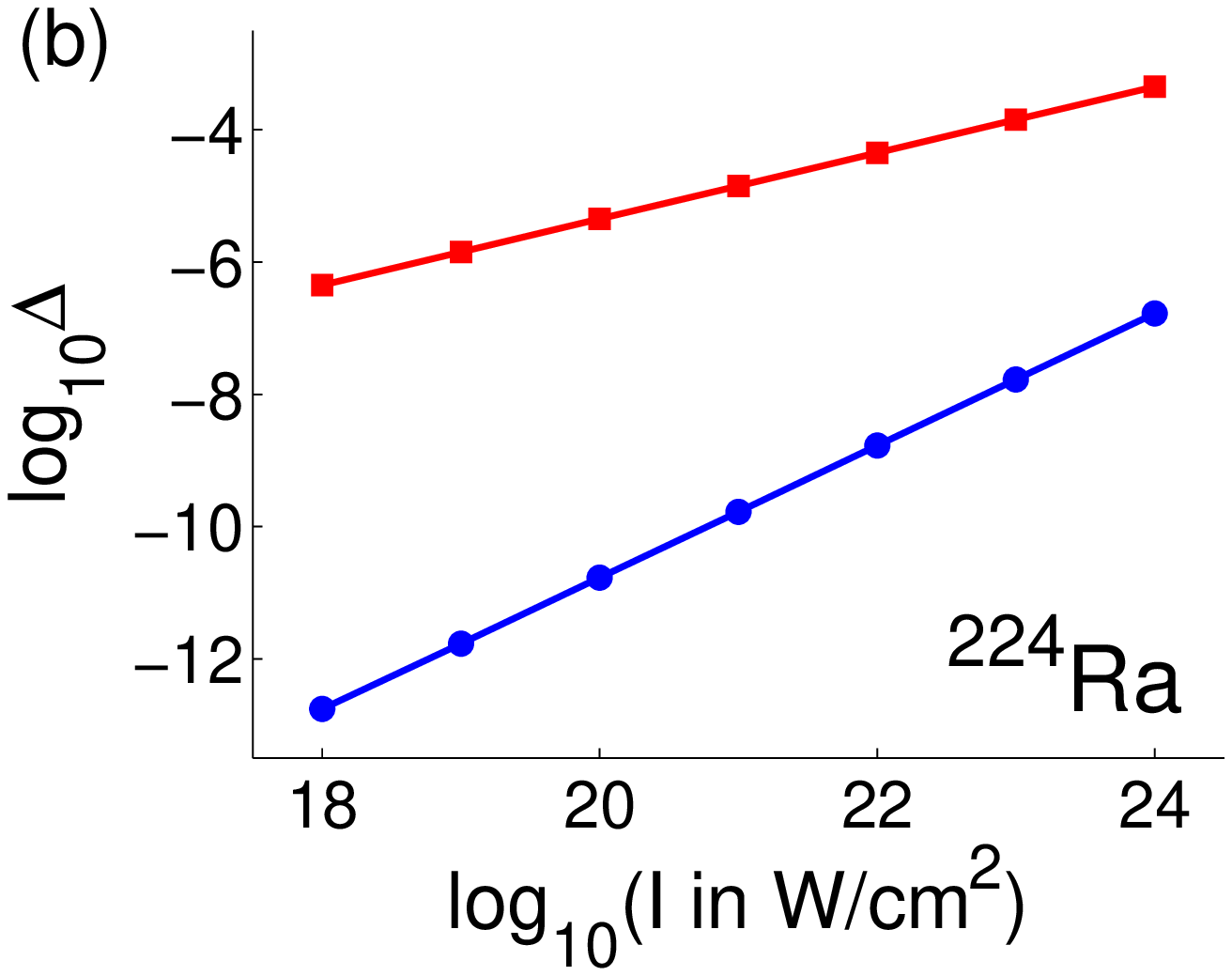}
\includegraphics[width=5cm]{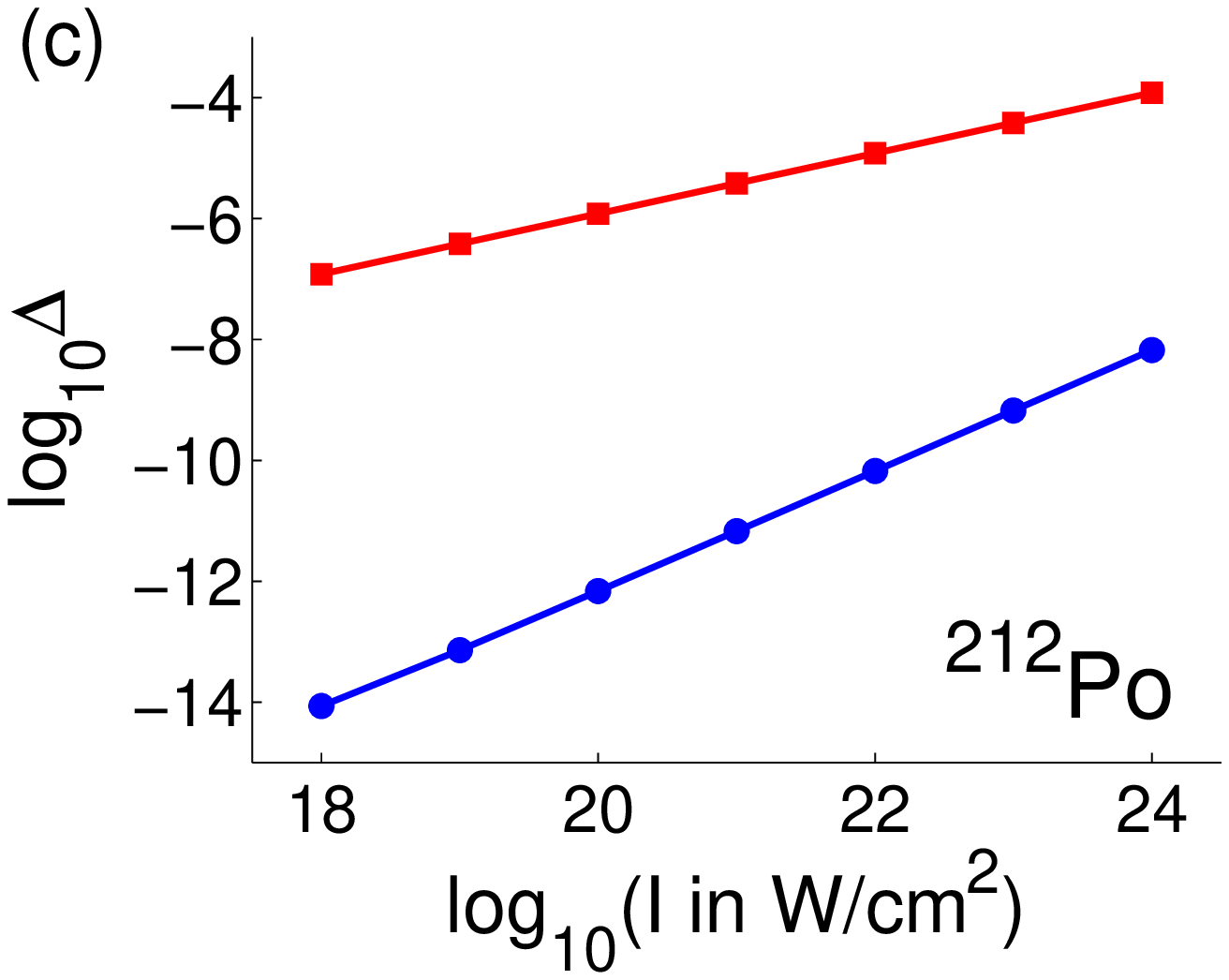} 
\caption{Angle-resolved (red squares) and angle-integrated (blue circles) modifications to the alpha decay penetrability, as a function of laser intensity. Both $\Delta$ and the intensity $I$ are plotted in the logarithmic scale, with which both curves are linear. The red curves have slope 0.5, due to a linear dependency on the laser field strength. The blue curves have slope 1.0, due to a quadratic dependency on the laser field strength.}  \label{f.0_total}
\end{figure}

Figure \ref{f.0_total} shows the angle-resolved (red squares) and angle-integrated modifications (blue circles) to the alpha decay penetrability. The relative modification $\Delta$ and the laser intensity $I$ are plotted in the logarithmic scale, therefore both curves are linear. The slope of the red curves is 0.5, due to a linear dependency on the laser electric field, while the slope of the blue curves is 1.0, due to a quadratic dependency on the laser electric field.

As analyzed in the previous subsection, for the intensities used in the current article, as well as laser intensities available in the near future, the response of the alpha decay to the external laser field is dominantly linear. Observing from a particular spatial angle, the modification to the alpha penetrability depends (almost) linearly on the laser field strength. The linear response cancels if the angle-integrated modifications are considered, leaving the quadratic response dominating.

\section{Conclusion}

We report in this article a combined theoretical and numerical study on the possible influences of strong laser fields on the nuclear alpha decay process. We use realistic and quantitative alpha-nucleus potentials and aim at obtaining quantitative evaluations of the laser influences.

We first show that the alpha penetrability (or equivalently the nuclear half life) can indeed be modified by strong laser fields to some small but finite extent, with laser intensities expected to be achievable in the forthcoming years especially with the under-constructing ELI facility. We also predict that alpha decays with lower decay energies are easier to be modified than those with higher decay energies, due to longer tunneling paths for the laser electric field to act on. This is a somewhat counterintuitive result.

We point out that compared to the alpha-nucleus potential, the additional laser potential is weak, even with the highest laser intensities achievable in the near future. The response of alpha decay to the laser field is shown to be restricted to the lowest two orders (linear and quadratic). Angle-resolved alpha penetrability is shown to be a first-order process, depending linearly on the laser field strength. Whereas angle-integrated alpha penetrability is shown to be a second-order process, depending quadratically on the laser field strength, or linearly on the laser field intensity. Future experiments investigating laser-modified alpha decay processes should start with angle-resolved observables.

\section*{Acknowledgement}

We acknowledge funding support from China Science Challenge Project No. TZ2018005, China NSF No. 11774323 and No.11725417, NSAF No. U1730449, National Key R\&D Program No. 2017YFA0403200.

\end{document}